\documentclass[amsmath,amssymb,aps,prx,notitlepage,superscriptaddress,longbibliography]{revtex4-2}
\usepackage{xcolor}
\usepackage{graphicx}
\usepackage{amsmath}
\usepackage{amssymb}
\usepackage{bbold}
\usepackage{bm}
\usepackage{hyperref}
\usepackage{comment}
\usepackage{array}

\newcommand{\cor}[1]{\textcolor{red}{#1}}

\begin{document}
\title{Optical estimation of unitary Gaussian processes without phase reference using Fock states}
\begin{abstract}
Since a general Gaussian process is phase-sensitive, a stable phase reference is required to take advantage of this feature. When the reference is missing, either due to the volatile nature of the measured sample or the measurement's technical limitations, the resulting process appears as random in phase. Under this condition, we consider two single-mode Gaussian processes, displacement and squeezing. 
We show that these two can be efficiently estimated using photon number states and photon number resolving detectors. 
For separate estimation of displacement and squeezing, the practical estimation errors for hundreds of probes' ensembles can saturate the Cram\'{e}r-Rao bound even for arbitrary small values of the estimated parameters and under realistic losses. The estimation of displacement with Fock states always outperforms estimation using Gaussian states with equivalent energy and optimal measurement. For estimation of squeezing, Fock states outperform Gaussian methods, but only when their energy is large enough.
Finally, we show that Fock states can also be used to estimate the displacement and the squeezing simultaneously.
\end{abstract}

\author{Changhun Oh}
\affiliation{Department of Physics and Astronomy, Seoul National University, Seoul 08826, Korea}
\affiliation{Pritzker School of Molecular Engineering, The University of Chicago, Chicago, Illinois 60637, USA}
\author{Kimin Park}
\affiliation{Department of Optics, Palack\'y University, 17. listopadu 1192/12, 771 46 Olomouc, Czech Republic}
\affiliation{Center for Macroscopic Quantum States (bigQ), Department of Physics, Technical University of Denmark, Building 307, Fysikvej, 2800 Kgs. Lyngby, Denmark}
\author{Radim Filip}
\affiliation{Department of Optics, Palack\'y University, 17. listopadu 1192/12, 771 46 Olomouc, Czech Republic}
\author{Hyunseok Jeong}
\affiliation{Department of Physics and Astronomy, Seoul National University, Seoul 08826, Korea}
\author{Petr Marek}
\affiliation{Department of Optics, Palack\'y University, 17. listopadu 1192/12, 771 46 Olomouc, Czech Republic}
\maketitle

\section{Introduction}
Quantum metrology with light estimates unknown parameters of quantum processes and reveals the limits of the existing measurements from treating the measuring probes as physical systems in specific optimized quantum states \cite{giovannetti2004quantum,giovannetti2006quantum,giovannetti2011advances, pirandola2018advances}. 
Optical interferometry is one the main examples of improvement brought in by using quantum states \cite{demkowicz2015quantum}. There, quantum noise entering the interferometer through the beam splitter's idle port can be reduced by using squeezed states of light \cite{caves1981quantum, olivares2009bayesian, berni2015ab, oh2019optimal}. The effect is robust enough to find a place in practical applications where high precision is required, such as detecting gravitational waves \cite{aasi2013enhanced}. A specific case of optical interferometry relies on homodyne detection \cite{schleich2011quantum}, where the reference arm of the interferometer is represented by a classical local oscillator beam \cite{olivares2009bayesian, berni2015ab, oh2019optimal}, which leaves only the probe to be prepared in a quantum state. Another approach takes advantage of controlling the full interferometer and preparing joint quantum NOON states in both of its arms \cite{lee2002quantum, dowling2008quantum, mccormick2019quantum}. The probes in quantum metrology are not limited to light. Atomic interferometry using collective states of atoms \cite{bongs2019taking, PhysRevLett.125.100402} and quantum optomechanics employing state of motion of a massive mirror \cite{qvarfort2018gravimetry, schneiter2020optimal} have both been studied for gravimetry. Similarly, quantum states of trapped ions \cite{dalvit2006quantum, wolf2019motional} can be applied towards the detection of weak electric fields.

One thing all of these methods have in common is the use of quantum states that are nonclassical \cite{tan2019nonclassical}, which means that they cannot be described by classical physics alone. The nonclassical states in optical systems can be further divided into Gaussian and non-Gaussian. Gaussian states, such as squeezed states \cite{caves1981quantum, olivares2009bayesian, berni2015ab, oh2019optimal} can be described by Gaussian functions in phase space, while the non-Gaussian cannot. One drawback shared by both kinds of nonclassical states is their vulnerability to imperfections. While the nonclassical quantum states of probes have been shown to significantly outperform the classical methods in ideal cases~\cite{olivares2009bayesian, mccormick2019quantum}, this improvement can vanish under realistic experimental conditions. In Gaussian scenarios \cite{genoni2013optimal, pinel2013quantum, vsafranek2015quantum, vsafranek2016optimal, nichols2018multiparameter, vsafranek2018estimation, oh2019optimal, oh2019optimal2}, such as employing squeezed states to boost interferometry, the losses of the quantum state simply reduce its effectiveness \cite{berni2015ab, oh2019optimal}. In non-Gaussian cases, such as employing so-called NOON states used to estimate phase, losses can completely remove the quantum advantage \cite{dowling2008quantum, escher2011general}. Even though recent progress tries to alleviate this effect \cite{slussarenko2017unconditional}, one question that needs to be addressed for all quantum metrology proposals is whether its benefits survive contact with practical reality.

Fluctuation of phase is one of the major sources of imperfection in optical interferometry \cite{genoni2011optical, genoni2012optical, escher2012quantum, vidrighin2014joint, szczykulska2017reaching, aguilar2020robust} and quantum communications \cite{fanizza2020classical, zhuang2020entanglement}. It can take the shape of random phase fluctuation in the sample, but also of the inability to lock the process in the sample to the phase of the probe. There are also situations in which there are no detectors to capitalize on the phase. This can happen in optics when broadband photon counting detectors are used instead of homodyne detection \cite{eckstein2011highly}.
However, control of the phase is not necessary for estimating the strength of the phase-sensitive operations, such as coherent displacement and nonclassical squeezing process.
It was shown in a recent trapped-ion experiment \cite{wolf2019motional}, where the size of displacement, a generally phase-sensitive operation, was estimated without any kind of phase reference, just relying on Fock state preparation and measurement.


In this paper, we extend this idea to optical experiments and show that optical estimation of the strength of unitary Gaussian operations, displacement and squeezing, can be indeed realized without any phase reference with Fock states and measurements on the photon number basis. For displacement, this approach surpasses even optimal Gaussian methods, which are based on homodyne detection and squeezed states with equivalent energy and which use the phase reference. In squeezing estimation, Fock states are generally comparable or even slightly inferior to Gaussian methods at low energies and overcoming them for larger energies of the probe states. Our probe states do not need to be pure and can have an advantage over the Gaussian probe states even for realistic losses of the order of 20\%. Finally, we show that this approach can be useful for the simultaneous estimation of both quantities \cite{chiribella2006joint}.



\section{Model of the process}
In our quantum sensing protocol for Gaussian processes without a stable phase reference, represented by a mixture of quantum evolution with all phases, we start with an ensemble of $M$ probes prepared in a well-defined quantum state that is fully under our control. 
The probes then sequentially interact with a sample, undergoing weak phase-randomized Gaussian evolution in the process, and are measured by a specific measurement. 
The measured data are then evaluated using maximum likelihood estimator (MLE) to extract the unknown parameters of the Gaussian operation. 
In our analysis we will focus on photon number resolving detectors (PNRD).


Action of a general single-mode Gaussian channel \cite{serafini2017quantum} can be represented by a linear transformation of quadrature operators of the field $\hat{x}$ and $\hat{p}$, with $[\hat{x},\hat{p}] = i$. If we arrange the quadratures into a vector $\hat{\xi} = (\hat{x},\hat{p})^\text{T}$, the general Gaussian operation transforms them into a new vector
\begin{equation}\label{}
    \hat{\xi}' = V \hat{\xi} + \alpha  + (\hat{x}_{E},\hat{p}_E)^\text{T},
\end{equation}
where $V$ is a real matrix with $|V|<1$, $\alpha = (\alpha_x,\alpha_p)^\text{T}$ is a vector of real values, and $\hat{x}_E$ and $\hat{p}_E$ are zero mean value Hermitian operators of the environment that satisfy  $[\hat{x}_E,\hat{p}_E] \ge i(1-|V|)$ and their statistics is Gaussian. Among these terms,  vector $\alpha$ models displacement, matrix $V$ includes phase shift, squeezing, and losses, and the pair of environment operators contributes added noise. In the absence of imperfections, the operation is unitary and is fully described by the vector $\alpha$ and matrix $V$ with $|V|=1$. 
Such operation can be decomposed into three separate processes. Phase shift, represented by operator $\hat{R}(\phi) = e^{-i\phi\hat{n}}$, displacement represented by $\hat{D}(\alpha) = \exp(\alpha \hat{a}^\dag - \alpha^*\hat{a})$ and squeezing represented by operator  $\hat{S}(\xi) = \exp[(\xi \hat{a}^{\dag 2} - \xi^*\hat{a}^2)/2]$. Here, $\hat{a} = (\hat{x} + i\hat{p})/\sqrt{2}$ denotes the annihilation operator of the field and $\hat{n} = \hat{a}^{\dag}\hat{a}$ is the photon number operator. Although these operations do not commute in general, the order of these operations can be arbitrary - what matters is their final product and different orderings can lead to the same overall operation if the parameters are suitably changed.

Let us now consider a scenario without a well-defined phase. 
The Gaussian effect on the mechanical probe can be decomposed by a sequence of displacement, phase shift, and squeezing, where the strengths of the displacement and squeezing are of main interest. In optical testing, it removes the need to lock the pump of the nonlinear process appearing in the sample, such as weakly nonlinear waveguide, to the phase of the signal. 
Both displacement and squeezing are phase-sensitive operations. 
If their phase cannot be locked to the signal or detectors are phase-insensitive, it is considered random. 
In the extreme case the phase is assumed to be uniformly distributed on the interval $[0,2\pi)$ by the principle of maximum entropy \cite{jaynes1968prior}. In any case, the operation is no longer unitary and needs to be expressed in terms of map which transforms the probe state $\hat{\rho}_{in}$ as 
 $\mathcal{M}$:
\begin{align}\label{evolution_map}
 &\hat{\rho}_f(N_c,N_s) = \mathcal{M}(\hat{\rho}_{in}) = \int_0^{2\pi} \frac{d\phi_1}{2\pi}  \int_0^{2\pi} \frac{d\phi_2}{2\pi}  \hat{D}(\sqrt{N_c}e^{i\phi_1})\hat{S}(\sqrt{N_s} e^{i\phi_2})\hat{\rho}_{in}\hat{S}^{\dag}(\sqrt{N_s} e^{i\phi_2}) \hat{D}^{\dag}(\sqrt{N_c}e^{i\phi_1}).
\end{align}
In contrast to a general pure Gaussian unitary operation with five parameters, the phase indeterminate operation has only two free parameters: $N_c$ related to the  average linearized energy  added by the displacement, and $N_s$  related to the energy added by the squeezing. On vacuum state $\hat{\rho}_{in}=|0\rangle\langle0|$, the added energies of the two operations are given as $N_c$ for displacement and $\sinh^2 (\sqrt{N_s})$ for squeezing. In the limit of weak strengths $N_s$ and $N_c$, which is the regime we are interested in, phase-insensitive displacement and squeezing operations commute and the average photon number is increased by $N_s+N_c$.

Two main effects that are not part of this model and yet play a significant role in practical situations are the fluctuations of parameters $N_S$ and $N_c$, and linear optical losses. 
Fluctuation of the parameters arises due to instability of the estimation process. In this case, what is usually estimated is the mean value of the parameters with their known fluctuation. The other effect, optical losses, naturally appear when part of information contained in the probe is lost. In the case of fast short-time passage through the sample, the majority of losses will appear either during the in-coupling and out-coupling of the probe. As a consequence, they can often be estimated separately and we can treat them as a known parameter in the estimation of the displacement and squeezing.
We will discuss both of these imperfections more later in the text.

Before we proceed to the analysis of the protocol, let us establish some theoretical framework by briefly recalling the quantum Cram\'{e}r-Rao (QCR) inequality and the quantum Fisher information (FI).
QCR inequality states that for a given probe and channel that encodes an unknown parameter of interest $\theta$, the estimation error of any unbiased estimator is bounded by the inverse of quantum FI \cite{helstrom1976quantum, braunstein1994statistical},
\begin{align}\label{crb}
\Delta^2\theta \geq \frac{1}{M H(\theta)},
\end{align}
where the variance $\Delta^2\theta=\langle (\theta^\text{est}-\theta)^2\rangle$ is the estimation error, and $M$ is the number of trials in an experiment, and $H(\theta)=\text{Tr}(\hat{\rho}_\theta \hat{L}_\theta^2)$ is the FI of the probe state after the encoding.
It is known that the lower bound is asymptotically saturable by using MLE \cite{fisher1925theory, braunstein1992large, braunstein1992quantum}.
In experiments, the optimal precision suggested by equality in (\ref{crb}) is obtained by an optimal positive operator valued measurement (POVM),
which can be found by the eigenbasis of the symmetric logarithmic derivative operator $\hat{L}_\theta$ satisfying an equation,
\begin{align}
\frac{\partial \hat{\rho}_\theta}{\partial \theta}=\frac{\hat{\rho}_\theta \hat{L}_\theta+\hat{L}_\theta\hat{\rho}_\theta }{2}.
\end{align}
If the density matrix of the output state is diagonalized as $\hat{\rho}_\theta=\sum_n\rho_n |\psi_n\rangle\langle\psi_n|$, the symmetric logarithmic derivative operator is given $\hat{L}_\theta=2\sum_{n,m}\frac{\langle\psi_n|\partial_\theta\hat{\rho}_\theta|\psi_m\rangle}{\rho_n+\rho_m} |\psi_n\rangle\langle\psi_m|$~\cite{paris2009quantum}.
When the optimal measurement is chosen, the classical FI
\begin{align}
F(\theta)=\sum_n \frac{1}{p(n|\theta)}\left(\frac{\partial p(n|\theta)}{\partial \theta}\right)^2
\end{align}
becomes the quantum FI where $p(n|\theta)=\text{Tr}(\hat{\rho}_\theta \hat{\Pi}_n)$ with $\{\hat{\Pi}_n\}$ being an optimal POVM, i.e. the projectors of eigenstates of $\hat{L}_\theta$.
According to QCR inequality \eqref{crb}, FI lower-bounds the estimation error obtainable during the actual measurement by $\Delta^2 N_c=1/M F$ where $M$ is the number of trials.
In a general scenario, however, there is no guarantee that this bound can be achieved with a practical number of $M$ although it is achievable using the MLE in an asymptotic regime of $M\rightarrow \infty$.
In the following sections, we present the estimation error of our scenario obtained by the MLE for a finite number of $M$.

Map (\ref{evolution_map}) is phase-insensitive; it commutes with any phase shift applied to the state of the probe. Consequently, if the probe state is phase-insensitive it remains so.
This suggests that a well-defined phase of the probe may not be required for optimal estimation. This can be illustrated on an example of probe prepared in Fock state $\hat{\rho}_{in}=|m\rangle\langle m|$. Such probe is pure but completely phase-insensitive. In the limit of weak strengths $N_c, N_s \ll 1$, map  (\ref{evolution_map}) transforms the initial pure state of the probe into a mixture of Fock states with weights as:
\begin{align} \label{eq:binary1}
p(m-2|N_c,N_s) &\simeq  N_s m(m-1)/4  \\
p(m-1|N_c,N_s) &\simeq  N_c m  \\
p(m|N_c,N_s) &\simeq   1 - N_c(2m+1) - N_s(m^2+m+1)/2  \\
p(m+1|N_c,N_s) &\simeq  N_c(m+1)  \\
p(m+2|N_c,N_s) &\simeq  N_s(m+2)(m+1)/4. \label{eq:binary2}
\end{align}
Since the state is diagonal in the Fock basis these weights can be perfectly measured by PNRD. In this important limit we can see that the displacement and squeezing unitaries act in a complementary way - displacement changes the photon number by one, squeezing changes it by two. This indicates that in this limit the two operations can be discerned independently.


The measured data, with the help of Eqs.~\eqref{eq:binary1}-\eqref{eq:binary2}, can be used to construct MLEs for each parameter. Let us denote the number of outcomes corresponding to detecting particular state $|k\rangle$ by $n_k$ with $k\in\{m-2,m-1,m,m+1,m+2\}$; thus, $\sum_{k=m-2}^{m+2} n_k=M$, where $M$ is again the total number of trials.
We simply ignore the outcome out of the above range since the probability is negligible for very small strength of the signal.
By maximizing the log-likelihood function $\log L(\mathbb{D}=\{n_k\}|N_c,N_s)=\sum_k n_k\log p(k|N_c,N_s)$ for each parameter $N_c$ and $N_s$, one can find that the MLE is written as
\begin{align}\label{weakML}
N_c^\text{est}=\frac{n_{m-1}+n_{m+1}}{M(2m+1)}, ~~~\text{and}~~~ N_s^\text{est}=\frac{2(n_{m-2}+n_{m+2})}{M(m^2+m+1)},
\end{align}
and that they are unbiased for any $M>0$, i.e., $\langle N_c^\text{est}\rangle=N_c$ and $\langle N_s^\text{est}\rangle=N_s$, where the bracket represents the average over all possible outcomes.
The form of the MLEs, in particular the exclusive use of the respective count numbers $n_k$, implies that each parameter can be estimated simultaneously without knowing the other parameter.

In addition, from the same equations we can derive classical FI to evaluate the estimation of $N_c$ and $N_s$ by using a PNRD, which we can then compared to the upper limit given by quantum FI for the optimal detector. For the case of phase-insensitive states, PNRD gives us full available information and the classical FI is equal to quantum FI. Based on Eqs.~\eqref{eq:binary1}-\eqref{eq:binary2}, the classical FI of Fock states can be approximately found to be:
\begin{equation}\label{disp_qfi}
   F(N_c)\approx\frac{2m+1}{N_c},
\end{equation}
for displacement and
\begin{equation}\label{sq_qfi}
   F(N_s)\approx \frac{m^2+m+1}{2N_s},
\end{equation}
for squeezing in the limit of $N_c, N_s \ll 1$.
In this limit, one can derive the average estimation error of ML estimators \eqref{weakML}
\begin{align}\label{weakVar}
\Delta^2 N_c\approx \frac{N_c}{M(2m+1)}~~~\text{and}~~~ \Delta^2 N_s\approx \frac{2N_s}{M(m^2+m+1)},
\end{align}
which is consistent with the FI.
For both of them the performance improves with the increased Fock number of the probe, linearly in the case of displacement while quadratically in the case of the squeezing.
The different scalings of Fisher information is ascribed by the fact that weak displacement and squeezing operations are single-photon and two-photon processes, respectively, as shown in Eqs.~\eqref{eq:binary1}-\eqref{eq:binary2}.
Both quantities diverge as the signal decreases, but the relative estimation errors, which are given as inverse of the Fisher information relative to the signal, $R = \frac{1}{F(N_i) N_i}$ with $i = c,s$, attain constant value.
In the next sections, we will  analyze how these values can be obtained with a realistic number of probes and how is the procedure affected by realistic processing and imperfections in comparison to results obtainable with Gaussian resources.

\section{Displacement estimation with Fock state probe}
Let us first analyze situations in which the displacement operation is the only relevant effect.
In this case, the operation is represented by map (\ref{evolution_map}) with the squeezing parameter $N_s = 0$.
It transforms the initial Fock state $|m\rangle\langle m|$ into a mixed state
\begin{align}\label{displacement_channel}
\hat{\rho}_f(N_c)=\int_0^{2\pi}\frac{d\phi}{2\pi}\hat{D}(\sqrt{N_c}e^{i\phi})|m\rangle\langle m|\hat{D}^\dagger(\sqrt{N_c}e^{i\phi})=\sum_{n=0}^{\infty}p(n|N_c)|n\rangle\langle n|,
\end{align}
where
\begin{align}
p(n|N_c)=|\langle n|\hat{D}(\sqrt{N_c})|m\rangle|^2=\frac{m!}{n!}e^{-N_c}N_c^{n-m} L_m^{(n-m)}(N_c)^2,
\end{align}
is the conditional probability to detect $n$ photons for a given $N_c$ and $m$ with $L_m^{n}(x)$ being associated Laguerre polynomials \cite{de1990properties}.
Since the final state is diagonal in Fock basis, it has equal quantum and classical FI for PNRD:
\begin{equation}\label{nc_fisher}
   F(N_c)=\sum_{n=0}^\infty \frac{1}{p(n|N_c)}\left(\frac{\partial p(n|N_c)}{\partial N_c} \right)^2
=\frac{2m+1}{N_c},
\end{equation}
which is exactly the value approached by the approximate relation (\ref{disp_qfi}). It implies that in the limit of weak strength $N_c\ll1$, the protocol of the MLE of Eq.~\eqref{weakML} is the optimal procedure.
The derivation of the FI is supplied in Appendix A.
It should be noted that the monotonous increase of FI with the energy of the state $m$ requires the measurement in the Fock basis.
If the measurement was replaced by measurement of the mean energy $\langle \hat{n} \rangle$, the size of displacement could still be inferred, but the error of the measurement increases with $m$ (see Appendix B for detail).

\begin{figure}[t!]
\includegraphics[width=500px]{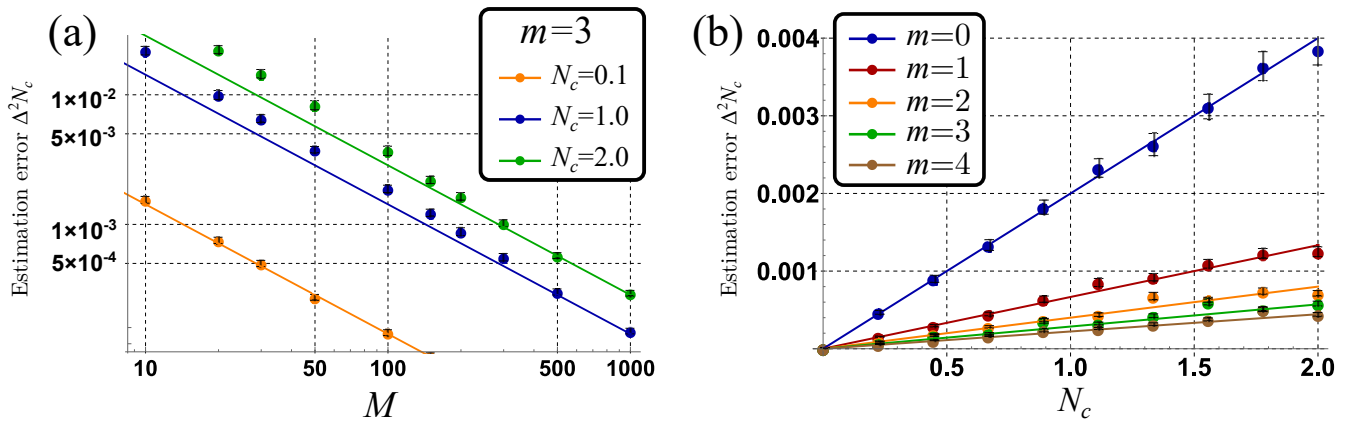}
\caption{Simulation for displacement estimation error $\Delta^2 N_c$.
(a) Against the number of copies $M$ for various strenths $N_c=0.1, 1.0, 2.0$ by a Fock state probe $|m=3\rangle$.
Lines and dots represent the inverse of FI (CR bound) and the estimation errors of simulations averaged over $3000$ trials using MLE. 
(b) Against the signal energy $N_c$ for various input probe Fock number $m=0,1,2,3,4$ with the number of copies $M=500$.
Again the lines and dots are inverse of FI and errors obtained by simulation averaged over $3000$ trials using MLE.
Throughout the paper, error bars represent twice the standard deviation of the obtained estimation error divided by the square root of the number of simulation runs.
}
\label{fig:dis1}
\end{figure}
%

To see whether practical errors can reach the bounds given by classical FI, we have performed a numerical simulation of the full protocol of estimating displacement with Fock state probes and PNRD.
For each scenario given by a different combination of $m$ and $N_c$, we have generated $3000$ sets of simulated data $\mathbb{D}$ with probability distributions $p(n|N_c)$  and evaluated them with a MLE, $N_c^{\text{est}}$, obtained by numerical maximization over a finite range, corresponding to a prior knowledge, of the log-likelihood function:
\begin{align}\label{likeli}
\log L(\mathbb{D}\equiv\{n_k\}|N_c)=\sum_{k=0}^\infty n_k \log p(k|N_c),
\end{align}
where $n_k$ is the number of outcomes for $k$ photons.
The estimated value $N_c^\text{est}$ was then compared to the true value $N_c$ to obtain the estimation error $\Delta^2 N_c = \langle(N_c^\text{est} - N_c)^2\rangle$ and compared to the QCR bound.
The results of the simulations can be seen in Fig.~\ref{fig:dis1}, where the simulated runs, marked by points, are compared to the bounds derived from quantum FI, represented by lines. In Fig.~\ref{fig:dis1}a we can see that, for probe in state $|3\rangle$, the realistic estimation error shows the same scaling as the bounds given by FI, saturates this bound already for $M = 500$ 
and that this scaling does not depend on the estimated value.
Both the dependence on $M$ and $N_c$ show that the the QCR bound is practically achievable with a finite $M$. Fig.~\ref{fig:dis1}b then confirms that this behavior holds even for probes prepared in different Fock states. It is worth noting that when the signal stength is small $N_c\ll1$, the process becomes a binary outcome estimation problem of $n_{m-1}+n_{m+1}$ and $n_m$ in Eq.~\eqref{weakML} so that the CR inequality is saturated by the ML estimator for any number of copies $M$ as shown in \eqref{weakVar}.

We emphasize that phase is not the only parameter that has a fluctuation in realistic experiment. 
Fluctuations of the strength of the displacement, which may arise from the fluctuation of an auxiliary pump, can also be considered. 
In this case, we can assume that the strength of the signal is a random variable following a normal distribution with a mean $N_c$ and a given variance depending on the amount of fluctuation and that the aim is to estimation the mean value $N_c$. We have numerically checked that in this case, the estimation error is additively increased by the amount of the fluctuation, which is shown in Appendix C. Keeping this in mind, from now on, we assume the fluctuation is small enough to neglect.

\begin{figure}[b]
\includegraphics[width=250px]{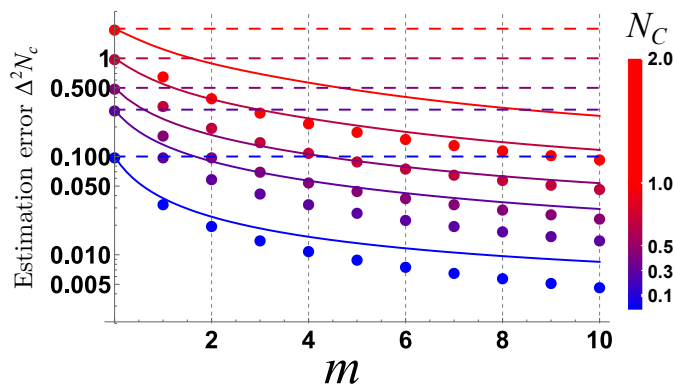}
\caption{Comparison between the lower bounds on estimation error given by the inverse of classical FI of Fock states with PNRD (dots), quantum FI of squeezed states with mean photon number equal to $m$  (solid lines), and quantum FI of coherent states (dashed lines) with mean photon number equal to $m$. Quantum  FI for squeezed states was calculated numerically. Different colors correspond to different values of the estimated parameter, $N_c=0.1, 0.3, 0.5, 1.0, 2.0$, and are specified by the color bar.
The squeezing required for energies equivalent to $m = 2,4,6,8,10$ amounts to 10.0,12.5, 14.1, 15.3, 16.2 dB.
}
\label{fig:dis_limit}
\end{figure}
Since quantum FI for Fock state probes is achievable by estimation with PNRD, we can use it for further analysis. For comparison we can consider practical Gaussian estimation methods employing Gaussian probes and stable phase between the input and the measured state. 
Notice that even if we assume stable phase for Gaussian probes and measurements, the phase of quantum operation of interest can be still random and the quantum operation is characterized by Eq.~\eqref{evolution_map}.

Even though the operation is random in phase, the stable phase is actually the most optimistic scenario for the Gaussian tools, because it enables noise reduction coming from squeezed vacuum fluctuations \cite{schafermeier2018deterministic}. If we lifted this assumptions and considered phase randomized Gaussian states or, equivalently, Gaussian states with phase randomized detection, their performance would be necessarily worse than that of Fock states. 
For our comparison we shall therefore consider quantum FI of phase-sensitive Gaussian states which might be higher than what can be achievable by Gaussian measurements. 
In this way we are comparing realistic estimation based on Fock states with the upper bound for Gaussian states.

Let us denote the output state from a channel with a parameter $\theta$ as $\hat{\rho}_\theta$.
In our case, the unknown parameter $\theta$ corresponds to the amount of energy $N_c$ pumped by the displacement.
The quantum FI of these density matrix can be found \cite{braunstein1994statistical}:
\begin{align}\label{eq:fisherfid}
H(\theta)=\frac{4[1-\mathcal{F}(\hat{\rho}_\theta,\hat{\rho}_{\theta+d\theta})]}{d\theta^2},
\end{align}
where $\mathcal{F}(\hat{\rho}_0,\hat{\rho}_1)=\left(\text{Tr}\sqrt{\hat{\rho}_0^{1/2}\hat{\rho}_1\hat{\rho}_0^{1/2}}\right)^2$ is the quantum fidelity between two quantum states $\hat{\rho}_0$ and $\hat{\rho}_1$.
Any Gaussian probe in a pure state can be expressed as a displaced squeezed vacuum state $\hat{D}(\beta)\hat{S}(\zeta)|0\rangle$.
Finding the optimal Gaussian probe requires maximization of the FI over the two parameters $\beta$ and $\zeta$ under the chosen constraints such as the total mean photon number in the input state.
One can easily check that the value of $\beta$ does not change the precision; thus, the optimal Gaussian probe is a squeezed state without any displacement possessing the lowest energy.
In Fig.~\ref{fig:dis_limit} we show the comparison of classical FI for Fock probes with PNRD, marked by dots, quantum FI of optimized Gaussian probes with equal energy, marked by solid lines, and quantum FI for vacuum state, marked by dashed lines, for the estimation of the unknown phase-insensitive displacement operations with various $N_c$.
We can see that the Fock state probes are superior to optimal Gaussian probes with the same mean photon number for the entire range of displacement strengths even though the former requires no phase stability and the latter may use arbitrary coherent detection schemes. This improvement is most prominent for large values $N_c$.
Note that since quantum FI is used to assess the achievable estimation precision of Gaussian states, and the final state from a squeezed state probe is generally phase-sensitive, we are implicitly assuming that a stable reference beam outside of the sensor is prepared and may be properly used for phase-sensitive measurement.
Without this reference the state needs to be treated as phase randomized squeezed state, which always performs worse than Fock state with equivalent energy, and is even definitely inferior to the vacuum state for low energies.


Let us now discuss the effects of optical imperfections, such as losses, to ensure the validity of the results in practical scenarios.
The photon-loss process, which is the main imperfection for light, can be described by quantum master equation  in the interaction picture as \cite{walls2007quantum}
\begin{align}\label{mastereq}
\frac{d\hat{\rho}}{dt}=\frac{\gamma}{2}\left(2\hat{a}\hat{\rho}\hat{a}^{\dag}-\hat{\rho}\hat{a}^\dagger\hat{a}-\hat{a}^\dagger\hat{a}\hat{\rho}\right),
\end{align}
where $\gamma$ is the loss parameter.
The loss rate is defined as $1-\eta=1-e^{-\gamma t}$ with $t\ge 0$ describing the monotonous decay of the coherence terms.
This dynamics can be equivalently described with   a virtual beam splitter interaction coupling the probe with a zero temperature bath. 
The losses can manifest either before the sample and thus represent the degradation of the probe, or after the sample and be related to imperfection of the measurement.
In the case of displacement estimation, the two losses act in almost the exactly the same way, only the one after the channel also reduces the measured quantity. To be more exact, if we represent the losses by completely positive trace-preserving map \cite{escher2011general}
\begin{equation}\label{decoherencemap}
    \mathcal{L}_{\eta}(\hat{\rho}) = \sum_{k = 0}^{\infty} \hat{A}_k \hat{\rho} \hat{A}_k^{\dag},\quad \hat{A}_k = \frac{\sqrt{1-\eta}^k}{\sqrt{k!}}\sqrt{\eta}^{\hat{a}^{\dag}\hat{a}}\hat{a}^k,
\end{equation}
then the losses after the displacing sample and before the displacing sample can be related as $\mathcal{L}_{\eta}[ \hat{D}(\sqrt{N_c})\hat{\rho}_{in} \hat{D}^{\dag}(\sqrt{N_c})] = \hat{D}(\sqrt{\eta N_c})\mathcal{L}_{\eta}(\hat{\rho}_{in} )\hat{D}^{\dag}(\sqrt{ \eta N_c})$. We can see that the losses affect the probe in exactly the same way and the only difference is in scaling of the estimated parameter. Displacement adding energy $N_c$ before the loss is equivalent to displacement adding energy $\eta N_c$ after the loss. In addition, this behavior remains also for the Gaussian states. For the sake of simplicity we can therefore consider only the losses before the sample. It should be noted that we assume $\eta$ is known through prior measurements and not a subject of the estimation.

\begin{figure}[t]
\includegraphics[width=480px]{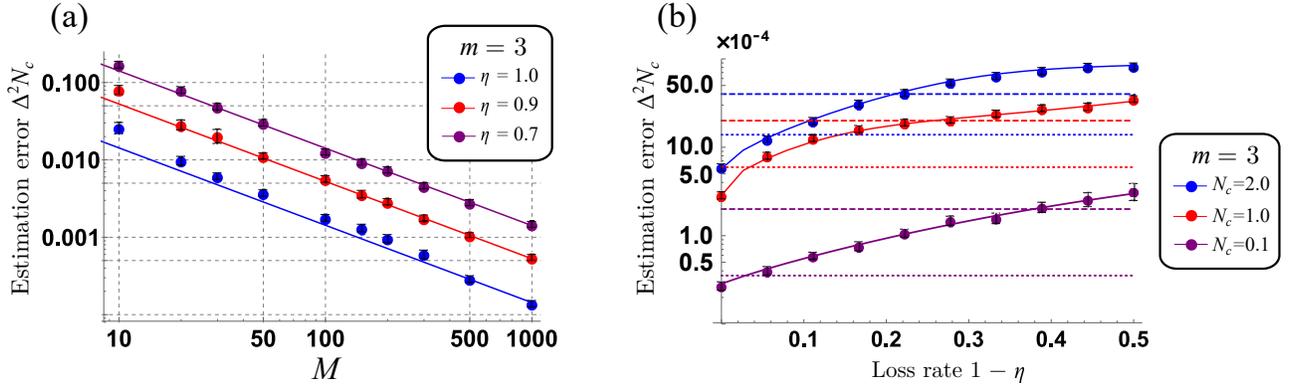}
\caption{Simulation for displacement estimation in the presence of photon-loss before the sample.
Dots and solid lines represent estimation errors and inverse quantum FI.
(a) Against the number of copies $M$ for different loss rates, $\eta = 1, 0.9, 0.7$ for $N_c = 1.0$ and $|m = 3\rangle$.
(b) Against the loss rate $1-\eta$ for  different values of the measured displacement $N_c=0.1, 1.0, 2.0$ with input probe $|m=3\rangle$ and $M=500$.
Classical limits (dashed lines) given by coherent states, and optimal Gaussian limits (dotted lines) by squeezed state with the same energy are evaluated from the inverse quantum FI without the losses.
}
\label{fig:dis_loss_limit}
\end{figure}
In numerical simulation shown in Fig.~\ref{fig:dis_loss_limit}, the theoretical Fock state distribution $p(n|N_c,\eta)$ was used to generate 1000 sets of data, which were then used, through MLE algorithm, to obtain the unknown value $N_c$ and estimate the error $\Delta^2 N_c$. The theoretical distribution $p(n|N_c,\eta)$ was obtained by using virtual beam splitter model represented by Eq.~\eqref{mastereq}. In Fig.~\ref{fig:dis_loss_limit}(a) are the numerically obtained errors for probe state $|3\rangle$ for three different levels of loss, represented by points, compared to quantum FI of the same probe states represented by solid lines. We can see that the errors  saturate the CR bound even in the presence of loss and keep the same scaling as the ideal scenario for various $M$. In Fig.~\ref{fig:dis_loss_limit}(b) are the same errors plotted with respect to range of losses and compared to optimal Gaussian states with the same mean photon number and without losses, represented by dashed lines, and to vacuum states, represented by dotted lines. We can see that even in the presence of losses, the estimation based on Fock states surpasses even the optimal methods using Gaussian states and optimal coherent measurements. We can also see different trends that appear for the comparison of Fock states to Gaussian and classical limits as the $N_c$ changes. As $N_c$ decreases, higher losses can be tolerated before the Fock state estimation falls behind the classical limit, or the shot-noise limit \cite{giovannetti2004quantum,giovannetti2006quantum,giovannetti2011advances, pirandola2018advances}, but at the same time lower losses are enough to fall behind methods using optimal Gaussian states. We have observed similar behavior for larger Fock states up to $m=7$ from a numerical calculation of Fisher information.



\section{Squeezing estimation with Fock state probe}
Let us now turn to the scenario in which we are interested only in the strength of an unknown squeezing operation that can be represented by map $\mathcal{M}$ in (\ref{evolution_map}) with $N_c = 0$.
We can analyze this scenario in the same way as the previous one. After this phase-insensitive squeezing operation, a Fock state $|m\rangle$ transforms to a mixed state
\begin{align}
\hat{\rho}_f(N_s)=\int_0^{2\pi}\frac{d\phi}{2\pi}\hat{S}(\sqrt{N_s}e^{i\phi})|m\rangle\langle m|\hat{S}^\dagger(\sqrt{N_s}e^{i\phi})=\sum_{n=0}^{\infty}p(n|N_s)|n\rangle\langle n|,
\end{align}
where \cite{kim1989properties}
\begin{align}
p(n|N_s)=|\langle n|\hat{S}(\sqrt{N_s})|m\rangle|^2&=\frac{n!m!}{2^{n-m}}\frac{\tanh^{n-m}\sqrt{N_s}}{\cosh^{2m+1}\sqrt{N_s}} S(\sqrt{N_s},m,n) ~~\text{ when } |m-n| \text{ is even} \\
&=0 ~~~~~~~~~~~~~~~~~~~~~~~~~~~~~~~~~~~~~~~~~~~~~~\text{ when } |m-n| \text{ is odd},
\end{align}
with
\begin{align}
S(r,m,n)=\left|\sum_k \frac{(-1)^k \sinh^{2k}r}{2^{2k}k!(m-2k)![k+(n-m)/2]!}\right|.
\end{align}
Here, the sum is taken for integers $k$ for which the argument of the factorials is positive.
For input Fock states we can explicitly calculate the quantum FI, which is again equal to classical FI for measurement in Fock state basis,
\begin{align}
F(N_s)=\frac{m^2+m+1}{2N_s}.
\end{align}
The derivation is similar to that of displacement estimation and is supplied in Appendix A.
We can thus analyze the estimation errors again in realistic scenario by numerical simulation.
Similar to the displacement estimation case, as the strength of the signal $N_s$ decreases, the estimation error by ML estimator of \eqref{weakML} saturates the QCR bound for any number of $M$ as shown in \eqref{weakVar}.

\begin{figure}[t]
\includegraphics[width=500px]{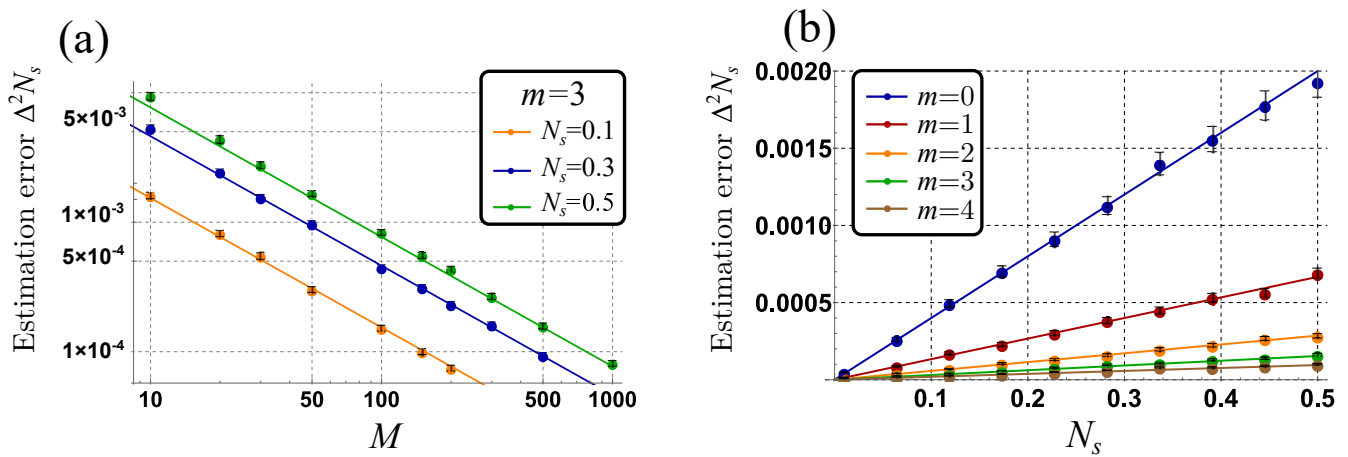}
\caption{Simulation for squeezing parameter estimation.
The solid lines and circle dots represent the estimation error based on CR bound and simulation averaged over 3000 trials using MLE with the full number statistics $p(n|N_s)$.
(a) Against the number of copies $M$ for $N_s=0.1, 0.3,0.5$.
(b) Against the energy added by squeezing $N_s$ for $m=0,1,2,3,4$.
The number of copies for simulation is  $M=500$.}
\label{fig:sq}
\end{figure}
In Fig. \ref{fig:sq}(a) and (b), these numerically obtained errors, marked by dots, are shown relative to the number of copies $M$ and value of the measured squeezing $N_s$, respectively. They are again compared to quantum FI represented by solid lines.
In both cases we can see that the estimation with Fock states and PNRD again approaches the precision predicted by the CR bound for a broad range of parameters.
Increasing the photon number $m$ of the Fock state probe leads to  better performance for the estimation of the phase-insensitive squeezing operation.
Again, we note that the fluctuation of $N_s$ can also be considered and the same behavior as displacement estimation is presented in Appendix C.

\begin{figure}[b]
\includegraphics[width=500px]{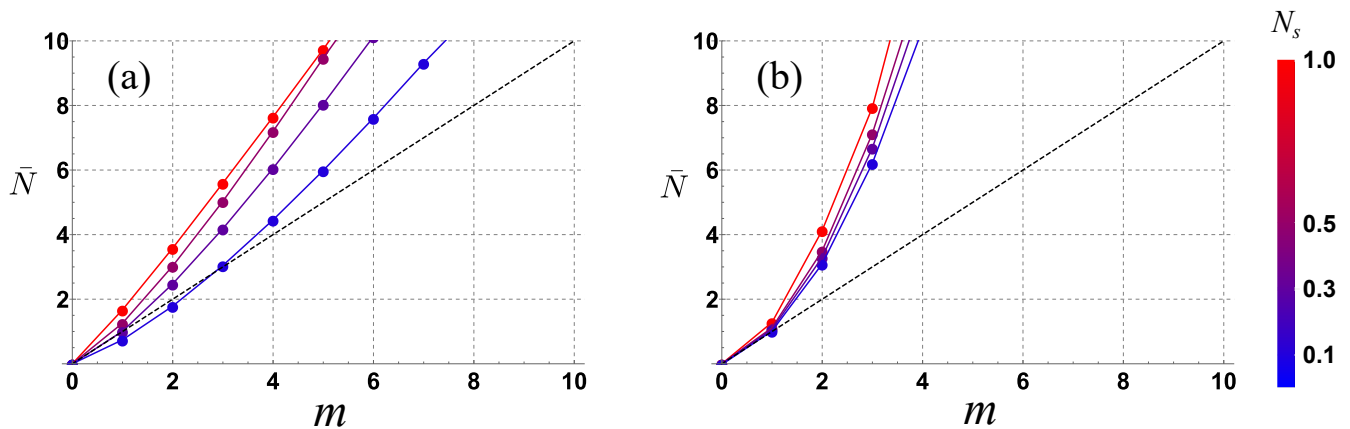}
\caption{Average photons required to attain the same quantum FI as classical FI of Fock states with PNRD (a) using squeezed states, (b) using coherent states. The left (right) plot shows that in overall, squeezed (coherent) states require more photon numbers than Fock states.}
\label{fig:sq_comp}
\end{figure}

Again we can compare the performance of Fock states to the Gaussian methods. When the Gaussian methods cannot take advantage of stable phase, the Gaussian probes can be expressed as mixtures of Fock states and therefore exhibit inferior performance. Numerical tests have confirmed that, in contrast to the displacement estimation, phase randomized squeezed states with arbitrary energy always perform worse than the vacuum state. 
To see the limits of the Fock-state-based estimation, we compare them to the optimal Gaussian estimation that takes advantage of phase reference. Here, in contrast to the displacement, a pre-displacement of the probe improves the estimation contrast while increasing mean photon number of the state of the probe. However, our numerical simulations revealed that the improvement gained by this displacement is, with regards to the number of photons added, smaller than what would be gained by additional squeezing with the same number of added photons.
We therefore compared the estimation error obtained by Fock states with the FI of Gaussian probes of coherent states or squeezed vacuum states without loss to obtain a strict threshold.
Fig.~\ref{fig:sq_comp} shows the mean photon number of (a) squeezed  or (b) coherent states that have the same estimation error as the inverse of the classical FI exhibited by various Fock states and PNRD.
The comparison shows that the Fock states are more energy-efficient than the Gaussian states in most cases. This is not the case for estimation of squeezing for small values of $N_s$ and small $m$. This is a rather interesting realization; metrology with Gaussian states can be applicable even in scenarios seemingly favouring the symmetry of Fock states.
It should be noted that in optical sensing, Gaussian states are generally easier to prepare than Fock states, but the difficulty varies wildly. Coherent states can be prepared routinely and are significantly more feasible than impure squeezed states. Preparing completely pure squeezed states, on the other hand, has difficulties comparable to preparation of Fock states. Comparison of equal mean photon numbers in Fig.~\ref{fig:sq_comp} shows that to attain the same precision with Gaussian states, significantly higher energy is required, which might be an issue for some applications \cite{taylor2016quantum}.
It is worthwhile to emphasize that our numerical calculation of quantum FI of Gaussian states showed by fitting with respect to the mean photon number that the scaling of quantum FI of coherent state and squeezed state is linear and quadratic with the mean photon number of the probe.

\begin{figure}[t]
\includegraphics[width=480px]{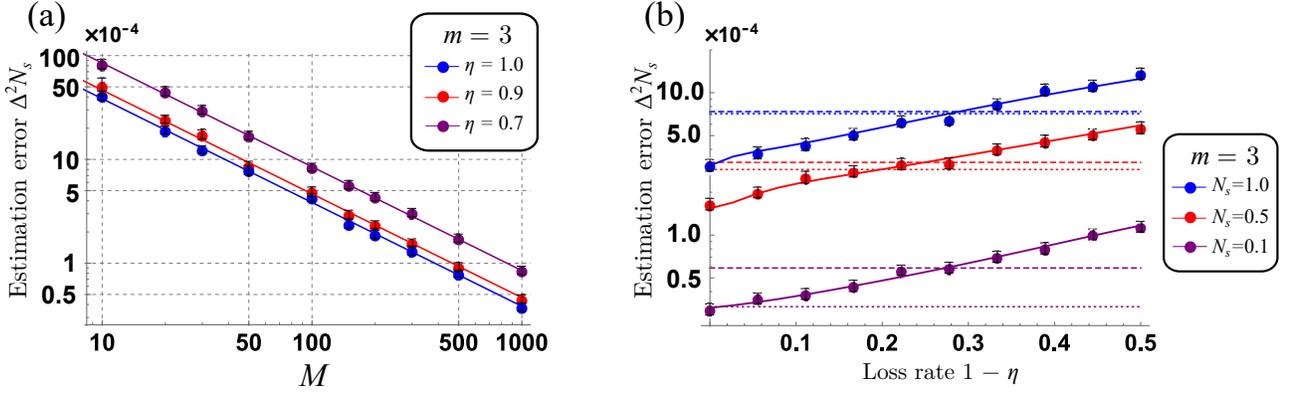}~~~~~~~~~~
\caption{Simulation for squeezing estimation with a Fock state $|m=3\rangle$ as an input probe in the presence of photon-loss.
The dots represent the estimation error based on simulation using MLE with the full number statistics $p(n|N_s)$, which are obtained by averaging over 1000 trials with $N_s=0.25$.
(a) Against the number of copies $M$.
Similarly to displacement estimation, the CR bound is saturated by MLE for a wide range of $M$ in the presence of photon-loss.
(b) Against loss rate $1-\eta$.
The dotted and dashed lines represent the estimation error limit of squeezed states and of coherent states without photon-loss, respectively. These classical estimation limits were evaluated as inverse of quantum FI for the states. The figure clearly showcases several scenarios in which Fock states subjected to loss provide better precision than pure Gaussian states.
 }
\label{fig:sq_loss}
\end{figure}
The bounds for coherent and squeezed states can now help us in evaluating the performance of the estimation with Fock states under losses.
The first important observation is that there is no simple relation in the estimation of squeezing  between the effect of losses before and after the sample.
This is because squeezing can lead to entanglement between the probe and the after-sample-bath, which then alters the properties of the probes.
However, in the limit of low values of estimated parameter $N_s \ll 1$, the effect of squeezing is linearized and this difference can be neglected.
In this regime, losses after the channel would alter the estimated value, but the qualitative behavior of the error rates would remain the same. Since this is the regime we are most often interested in, we can again, for the sake of simplicity of analysis, consider only the case with losses before the channel.

We again performed numerical simulation for the estimation errors of squeezing under loss,  which are shown in Fig.~\ref{fig:sq_loss} (a) relative to number of copies $M$ and  (b) relative to the loss rate $1-\eta$.
Both figures show that, similarly to the scenario of displacement estimation, the estimation errors under loss approach the CR bound and that the scaling with M remains consistent.
While for small values of $N_s$ and $m$ the Fock states and squeezed states with optimal coherent detection are comparable such as $N_s = 0.1$ and $m=3$, the Fock states enable attaining a better scaling of precision for large $N_s$ and $m$ as suggested by Fig.~\ref{fig:sq_comp} (b) and  Fig.~\ref{fig:sq_loss} (b) for small $1-\eta$.

\section{Simultaneous estimation of displacement and squeezing}
Finally, let us consider a general scenario in which both quantities, $N_c$ and $N_s$, appear at the same time and are estimated simultaneously.
This can be part of characterization of a general Gaussian process.
Another way this scenario can arise is when squeezing, the nonlinear process we want to characterize, is accompanied by noise with Poissonian distribution that can not be separated from the process. In this scenario we need to attempt simultaneous estimation of both quantities even though we are only ultimately interested in one. 

After the general channel (\ref{evolution_map}), the Fock state probe will be transformed to
\begin{align}\label{simul_map}
    \hat{\rho}_f(N_c,N_s) = \mathcal{M}(\hat{\rho}_{in}) 
=\sum_{n,k=0}^\infty w(n|k)q(k|m)|n\rangle\langle n|=\sum_{n=0}^\infty p(n|m) |n\rangle\langle n|,
\end{align}
where $q(k|m)=|\langle k| \hat{S}(r)|m\rangle|^2$, $w(n|k)=|\langle n|\hat{D}(\alpha)|k\rangle|^2$, and $p(n|m)=\sum_{k=0}^\infty w(n|k)q(k|m)$.
The lower bound of error on simultaneous estimation of $N_c$ and $N_s$, or multiparameter CR bound, is given by the quantum FI matrix
$C\geq H^{-1}$ \cite{helstrom1976quantum, szczykulska2016multi, gessner2018sensitivity, liu2019quantum} with covariance matrix $C_{N_c,N_c}=\langle (N_c^\text{est}-N_c)^2\rangle, C_{N_s,N_s}=\langle (N_s^\text{est}-N_s)^2\rangle, C_{N_c,N_s}=C_{N_s,N_c}=\langle (N_c^\text{est}-N_c)(N_s^\text{est}-N_s)\rangle$.
Here the matrix inequality $A\geq B$ means that $A-B$ is a positive semi-definite matrix.
Since the final state is always diagonal in the Fock basis, the quantum FI matrix, which is the same as the classical FI matrix based on the PNRD, can be written as
\begin{align}
H(N_c,N_s)=
\begin{pmatrix}
H_{N_c,N_c} & H_{N_c,N_s} \\
H_{N_s,N_c} & H_{N_s,N_s}
\end{pmatrix},
\end{align}
with
\begin{align}
H_{x,y}&=\sum_{n=0}^\infty \frac{1}{p(n|m)}\left(\frac{\partial p(n|m)}{\partial x}\right)\left(\frac{\partial p(n|m)}{\partial y}\right), \\
\end{align}
where $x,y$ are in $\{N_c, N_s\}$.
The classical multiparameter CR bound can also be asymptotically saturated by ML estimator.
From the multiparameter CR bound, we can extract the estimation errors of each parameter,
\begin{align}
\Delta^2 N_c&\geq \frac{H_{N_s,N_s}}{H_{N_c,N_c}H_{N_s,N_s}-H_{N_c,N_s}^2}=H_{N_c,N_c}^{-1}\left(1-H_{N_c,N_s}^2/H_{N_c,N_c}H_{N_s,N_s}\right)^{-1},~~~ \label{mcrb1}\\
\Delta^2 N_s&\geq \frac{H_{N_c,N_c}}{H_{N_c,N_c}H_{N_s,N_s}-H_{N_c,N_s}^2}=H_{N_s,N_s}^{-1}\left(1-H_{N_c,N_s}^2/H_{N_c,N_c}H_{N_s,N_s}\right)^{-1}.\label{mcrb2}
\end{align}

When more than one parameter in the process are involved, two main difficulties arise that may degrade the estimation error \cite{helstrom1976quantum, proctor2018multiparameter}.
First of all, as shown in inequalities ~\eqref{mcrb1} and \eqref{mcrb2}, the off-diagonal elements of FI matrix decrease the estimation error for fixed diagonal elements.
Non-vanishing off-diagonal elements of FI matrix imply that the parameters interplay each other in the process, so that one needs to know the other parameters in order to estimate a parameter of interest precisely.
On the other hand, when the off-diagonal element of the FI matrix vanishes, the estimation errors reduce to
\begin{align}
\Delta^2 N_c&\geq H_{N_c,N_c}^{-1},~~~ \label{mcrb_approx1}\\
\Delta^2 N_s&\geq H_{N_s,N_s}^{-1}.\label{mcrb_approx2}
\end{align}
Thus, when the off-diagonal element of the FI matrix is much smaller than the diagonal elements, the estimation error of each parameter is bounded by the inverse of each diagonal element of the FI matrix.
In this case, we can interpret the inequalities as that of a single-parameter estimation where any information about the other parameters is not required to estimate the parameter of interest.

The second difficulty is that even when one is estimating a single parameter, since the other parameters are involved, the diagonal elements of quantum (classical) FI may decrease.
For instance, when we estimate the squeezing parameter $N_s$, the displacement process in Map \eqref{simul_map}, written as $w(n|k)$, plays a role of a noisy process in the measurement setup.
Similarly when we estimate the displacement parameter $N_c$, the squeezing process in Map \eqref{simul_map}, written as $q(k|m)$, plays a role as a preparation error.
Thus, generally when more than one parameter is involved, the estimation error may decrease.
We investigate our case by numerical simulation focusing on these difficulties.


\begin{figure}[t]
\includegraphics[width=480px]{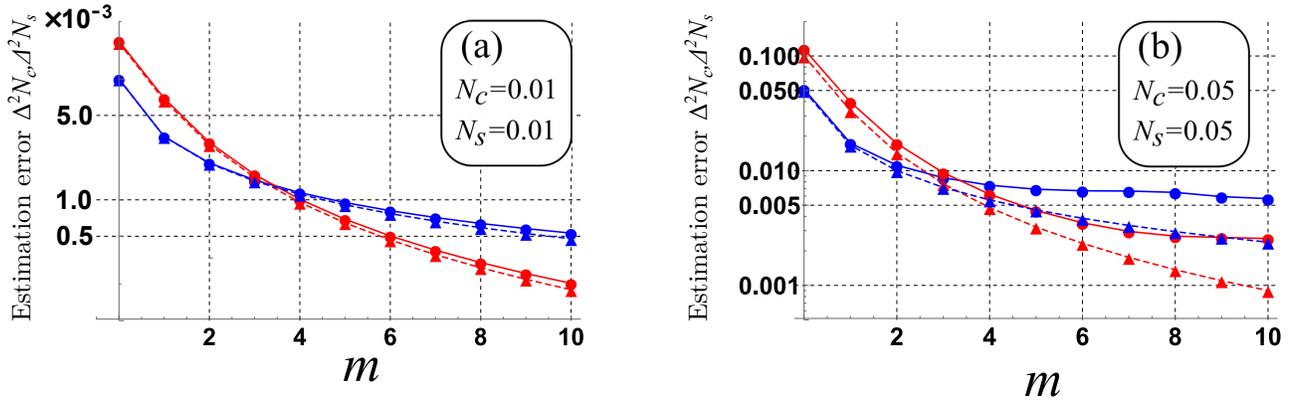}~~~~~~~~~~~~~
\caption{Simultaneous estimation errors for $N_c$ (blue  lines) and $N_s$ (red lines).
More specifically, solid lines with circles represent the lower bound of the esimation error in Eq.~\eqref{mcrb1} and Eq.~\eqref{mcrb2}, respectively.
Dashed lines with triangles represent the quantum FI of Fock state probe for estimating $N_c$($N_s$) when $N_s$($N_c$) is $0$ and known.
(a) shows values for estimating $N_c = N_s = 0.01$, (b) shows values for estimating $N_c = N_s  =0.05$.
}
\label{sim1}
\end{figure}

In Fig.~\ref{sim1}(a) and (b), we show the numerically calculated the lower bounds of the estimation errors for two different sets of measured values.
We numerically confirmed that the off-diagonal elements are very small compared to the diagonal elements ($H_{N_c,N_s}^2/H_{N_c,N_c}H_{N_s,N_s}<10^{-3}$ in Fig.~ \ref{sim1}(a) and $H_{N_c,N_s}^2/H_{N_c,N_c}H_{N_s,N_s}<1.2\times10^{-2}$ in Fig.~ \ref{sim1}(b)), which means that we do not suffer from the first difficulty in this regime and that the inverse of the diagonal components of FI matrix approximately give the lower bounds of the estimation errors of each quantity as written in Eqs.~\eqref{mcrb_approx1} and \eqref{mcrb_approx2}.
This is best seen for small values of estimated parameters in Fig.~\ref{sim1}(a) where the off-diagonal elements are truly negligible and errors are practically identical, whether both quantities are measured or just single ones.
The reason why simultaneous estimation for small values of estimated paramters works well is that when the estimated parameters are small, displacement and squeezing operations do not interfere each other because they are single-photon and two-photon processes, respectively, as emphasized in Eqs.~\eqref{eq:binary1}-\eqref{eq:binary2}.
In Fig.~\ref{sim1}(b), where the estimated parameters are not as small, the quantities start disturbing each other and the deviation from the individual estimations increase.
When the estimated parameters are not small enough, the role of the other parameter as a noisy process becomes so dominant that highly nonclassical states such as Fock states of large photon numbers may cease to give a small estimation error as shown in Fig.~\ref{sim1}(b).


\section{Conclusion}
In this work, we have investigated the possibility of using Fock states and photon number resolving detectors for parameter estimation of single mode Gaussian unitaries in the absence of stable phase reference.
This scenario is relevant in optical sensing when stable phase reference is unavailable through the sample, for example because of random nature of the examined operation and the light emitted by the sample is weak.

To accommodate both points of view, we have evaluated the performance of Fock states under realistic environment and compared them to the optimal performance of phase-sensitive Gaussian states with no loss and optimal quantum measurement. We found out that for estimation of both weak displacement and squeezing, the Fock states together with Fock basis detection can, already for ensemble of 500 trials saturate the Cram\'{e}r-Rao bound and provide error rates surpassing optimal Gaussian states with equivalent mean photon number.
Loss incurred in the sample or during preparation of the probes limits the quality of the estimation. The influence generally depends on the strength of the measured interaction. When $N_c,N_s \approx 0.5$, Fock states outperform the Gaussian bounds even when affected by 20\% losses. When $N_c,N_s\approx 0.1$, loss of 20\% can be tolerated when compared to coherent states, but less then 5\% loss brings the Fock states above the level of pure squeezed states. Interestingly enough, when estimating squeezing with low energy probes, Gaussian squeezed states surpass the Fock states even though the symmetry of the operation favors them.
 Simultaneous estimation of both squeezing and displacement is also possible without being disturbed from each other and it works best in the limit of small parameters, $N_c,N_s <0.1$, when the operations are effectively independent. Together, these features can allow either multi-parameter estimation of an optical Gaussian process in various systems, including atomic physics and solid-state physics, or estimation of new squeezing processes under inherent Poissoninan noise. The method can also be extended for estimation of higher order processes which encompass joint $n$-photon effects.

Experimental application of the procedure relies on Fock states and measurements in Fock basis. The measurement requires photon number resolving detectors. Transition edge sensors (TES) \cite{lita2008counting, calkins2013high, marsili2013detecting, harder2016single, burenkov2017full} is well known to be promising in this area as they are already capable of resolving up to 12 photons with estimated 0.98 detection efficiency \cite{sperling2017detector}. Alternatively, detector with photon number resolving capability can be constructed from an array of on-off detectors \cite{achilles2006direct, avenhaus2008photon, usuga2010noise, yukawa2013generating, harder2016local, yukawa2013generating, cooper2013experimental, bohmann2018incomplete}, or it can be, for purposes of proof-of-concept tests, replaced altogether by homodyne tomography. The photon number resolving detectors can be also used for preparation of the Fock states for the probes. Detecting a specific Fock state in one mode of a two-mode squeezed state generated by Optical Parametric Oscillator (OPO) projects the other mode into the same Fock state and is a technique often employed in quantum optics. It is also possible to generate the necessary Fock states by merging single photon states \cite{motes2016efficient}, which can be generated by quantum dots \cite{bulgarini2014nanowire, ding2016demand, senellart2017high, dusanowski2019near, ollivier2020reproducibility}. A proof-of-principle experimental test of the estimation method could be immediately realized with Fock states $|1\rangle$ or $|2\rangle$, conditionally obtained from an OPO, measured by TES and homodyne tomography for the verification purposes.
Coherent displacement can appear by a weak crosstalk to a different mode occupied by a coherent state in an optical system \cite{jeong2014generation}, using the optomechanical coupling in an optomechanical system \cite{sekatski2014macroscopic}. Finally, the detection method can be also considered outside the area of quantum optics.  For example, 
Fock states were already employed for estimation of displacement and can be considered for estimation of squeezing on the same platform \cite{ge2019trapped, drechsler2020state}.




\section*{Acknowledgement}
We acknowledge project 19-19722J of the Grant Agency of Czech Republic (GA\v{C}R).
C. O. acknowledges support from NSF (OMA-1936118).
K. P. acknowledges Danish National Research Foundation through the Center of Excellence for Macroscopic Quantum States (bigQ, DNRF142).
R. F. also acknowledges the MEYS of the Czech Republic (grant agreements $02.1.01/0.0/0.0/16\_026/0008460$ and 8C20002) and the funding from European Union's Horizon 2020 (2014-2020) research and innovation framework programme under grant agreement No 731473 (ShoQC). Project ShoQC has received funding from the QuantERA ERA-NET Cofund in Quantum Technologies implemented within the European Union's Horizon 2020 Programme.
H. J. acknowledges the National Research Foundation of Korea through grants funded by the Korea Government (NRF-2018K2A9A1A06069933, NRF-2019M3E4A1080074 and NRF-2020R1A2C1008609).



\section*{Appendix A: quantum Fisher information}
\setcounter{equation}{0}
\renewcommand{\theequation}{A\arabic{equation}}
Let us consider a unitary operation $\hat{U}=e^{-i \hat{H} \theta}$ where the Hamiltonian $\hat{H}$ generates $\theta$.
In the case of displacement, $\hat{H}=\hat{a}+\hat{a}^\dagger$, and for squeezing, $\hat{H}=i(\hat{a}^2-\hat{a}^{\dagger2}$).
After the phase-randomized displacement or squeezing operation, the output state can be written as
\begin{align}
\hat{\rho}_f=\sum_{n=0}^\infty p(n|m)|n\rangle\langle n|=\sum_{n=0}^\infty |\langle n|\hat{U}|m\rangle|^2|n\rangle\langle n|.
\end{align}
Let us first consider displacement estimation, $\hat{U}=e^{\alpha \hat{a}^\dagger-\alpha^* \hat{a}}$.
Since the final state is diagonal in Fock basis, the derivation of classical Fisher information of $p(n|m)$ is sufficient.
The classical Fisher information for $|\alpha|$ is written as
\begin{align}
F(|\alpha|)=\sum_{n=0}^\infty \frac{1}{p(n|m)}\left(\frac{\partial p(n|m)}{\partial |\alpha|} \right)^2.
\end{align}
Assuming $\alpha$ to be real without loss of generality, the differential term is simplified as
\begin{align}
\left(\frac{\partial p(n|m)}{\partial |\alpha|} \right)^2&=\left(-\langle n|\hat{D}(\alpha)|m\rangle\langle m|\hat{D}^\dagger(\alpha)(\hat{a}^\dagger-\hat{a})|n\rangle +\langle n|\hat{D}(\alpha)(\hat{a}^\dagger-\hat{a})|m\rangle\langle m|\hat{D}^\dagger(\alpha)|n\rangle \right)^2 \\
&=\left(\langle n|\hat{D}(\alpha)|m\rangle\langle m|\hat{D}^\dagger(\alpha)(\hat{a}^\dagger-\hat{a})|n\rangle\right)^2+\left(\langle n|\hat{D}(\alpha)(\hat{a}^\dagger-\hat{a})|m\rangle\langle m|\hat{D}^\dagger(\alpha)|n\rangle\right)^2 \nonumber \\
&-2\langle n|\hat{D}(\alpha)|m\rangle\langle m|\hat{D}^\dagger(\alpha)(\hat{a}^\dagger-\hat{a})|n\rangle\langle n|\hat{D}(\alpha)(\hat{a}^\dagger-\hat{a})|m\rangle\langle m|\hat{D}^\dagger(\alpha)|n\rangle \\
&=4p(n|m)\langle m|\hat{D}^\dagger(\alpha)(\hat{a}^\dagger-\hat{a})|n\rangle\langle n|\hat{D}(\alpha)(\hat{a}-\hat{a}^\dagger)|m\rangle.
\end{align}
Finally, we obtain the Fisher information
\begin{align}
F(|\alpha|)=4\langle m|\hat{D}^\dagger(\alpha)(\hat{a}^\dagger-\hat{a})\hat{D}(\alpha)(\hat{a}-\hat{a}^\dagger)|m\rangle=4\langle m|(\hat{a}^\dagger-\hat{a})(\hat{a}-\hat{a}^\dagger)|m\rangle=8m+4.
\end{align}
Simiarly, one can check that $F(r)=2(m^2+m+1)$.

Since we are interested in the Fisher information about $N_c=|\alpha|^2$, one can use the chain-rule for the classical Fisher information such as,
\begin{align}
F(N_c)=\sum_{n=0}^\infty \frac{1}{p(n|m)}\left(\frac{\partial p(n|m)}{\partial N_c} \right)^2=\frac{1}{4N_c}\sum_{n=0}^\infty \frac{1}{p(n|m)}\left(\frac{\partial p(n|m)}{\partial |\alpha|} \right)^2=\frac{F(|\alpha|)}{4N_c}=\frac{2m+1}{N_c}.
\end{align}
Similarly, the Fisher information about $N_s=r^2$ is obtained by
\begin{align}
F(N_s)=\sum_{n=0}^\infty \frac{1}{p(n|m)}\left(\frac{\partial p(n|m)}{\partial N_s} \right)^2=\frac{1}{4N_s}\sum_{n=0}^\infty \frac{1}{p(n|m)}\left(\frac{\partial p(n|m)}{\partial r} \right)^2=\frac{F(r)}{4N_s}=\frac{m^2+m+1}{2N_s}.
\end{align}

\section*{Appendix B: Estimating displacement by mean intensity measurement}
\setcounter{equation}{0}
\renewcommand{\theequation}{B\arabic{equation}}
The advantage of the Fock state probes requires the Fock state measurement to be fully utilized. This can be shown by considering a different, more limited, measurement that measures only the average moments instead of the full Fock state distribution. This is the case of coarse grained intensity detectors which average over the individual results and allow acces only to moments
\begin{align}
\langle\hat{n}\rangle&=m+N_c \\
\langle\hat{n}^2\rangle&=m^2+2N_c(2m+1)+N_c^2 \\
\Delta^2 \hat{n}&=\langle\hat{n}^2\rangle-\langle\hat{n}\rangle^2=2N_c(m+1).
\end{align}
The signal-to-noise ratio (SNR) can be defined where the signal is the change in  the average photon number of the probe and the noise is the uncertainty in the photon number  as
\begin{align}
\text{SNR}=\frac{\langle \hat{n}\rangle-m}{\sqrt{\Delta^2 \hat{n}}}=\sqrt{\frac{N_c}{2(m+1)}}.
\end{align}
Here, we notice that SNR has completely inverse tendencies from the Fisher information against $m$ and $N_c$, i.e. a larger precision for for a small $m$ and a large $N_c$.
The linearized sensitivity $(\Delta^2 N_c)_{\hat{O}}=\Delta^2\hat{O}/|\partial\langle\hat{O}\rangle/\partial N_c|^2$ based on $\langle\hat{n}\rangle$ gives an information about the uncertainty of $N_c$ by measuring the observable $\hat{O}$ which is given as
\begin{align}
(\Delta^2 N_c)_{\hat{n}}=\frac{\Delta^2 \hat{n}}{\left(\partial \langle \hat{n}\rangle /\partial N_c \right)^2}=2N_c (m+1).
\end{align}
Similarly, we can test the linearized sensitivity based on the second moment,
\begin{align}
\langle\hat{n}^4\rangle-\langle\hat{n}^2\rangle^2&=2(4 m+1) N_c^3+\left(18 m^2+2 m+3\right) N_c^2+\left(8 m^3+2 m^2+6 m+1\right) N_c\simeq 8m^3 N_c, \\
\frac{\langle\hat{n}^4\rangle-\langle\hat{n}^2\rangle^2}{(\partial\langle\hat{n}^2\rangle/\partial N_c)^2}&\simeq \frac{1}{2}mN_c,
\end{align}
where the approximation is under the condition of a small $N_c$ and a large $m$.
It shows that the error based on the second moment also increases with $m$.
Note that a high linearized sensitivity indicates a less precise measurement, or a large error.
The error increases with $m$ and  the best probe  is the vacuum probe, and therefore the Fock state probes do not give any advantage.
Therefore, we know that the gain in Fisher information of Eq.~\eqref{nc_fisher} cannot be obtained by simply using the information about average photon number $\langle \hat{n}\rangle$.
These results imply that the information about mean values does not provide the scalings of sensitivity available by estimation strategies with PNRD.

\section*{Appendix C: Fluctuations of signal}
\begin{figure}[t]
\includegraphics[width=400px]{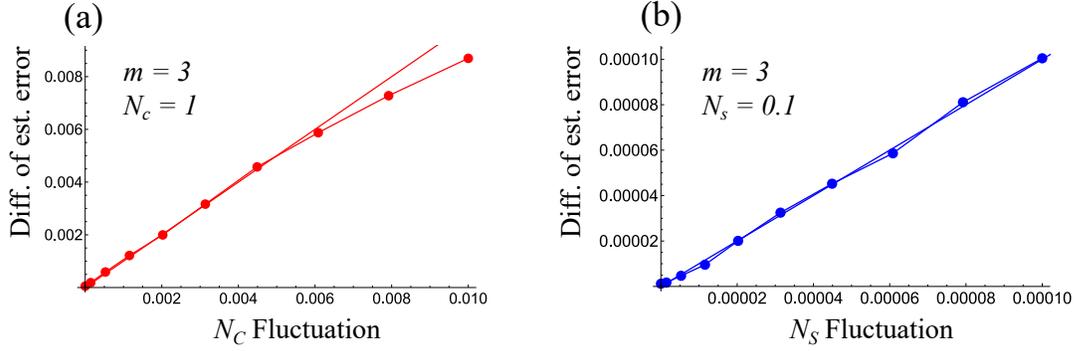}~~~~~~~~~~~~~
\caption{Additional estimation error that arises from fluctuations of the signals of (a) displacement and (b) squeezing. The additional estimation errors are the same as the fluctuations of the signals (solid lines) when fluctuations are small. We have averaged over 1000 trials and used $M=500$ copies.
}
\label{fluc}
\end{figure}

\setcounter{equation}{0}
\renewcommand{\theequation}{C\arabic{equation}}
In realistic experiments, fluctuations of pumps lead to a fluctuations of signals. Or, the signal itself may have a fluctuation. In this case, a natural approach is to estimate the mean value of the signal.
In this section, we analyze the effect of the fluctuation on the estimation error.
We have simulated the estimation of the strength of the displacement and squeezing, $N_c=1$ and $N_s=0.1$, with a Fock state $|m=3\rangle$.
In the Monte-Carlo simulation, each individual probe passes through a channel with different parameters, displacement and squeezing, and these fluctuate according a normal distribution with the mean $N_c$, $N_s$, respectively.
Fig. \ref{fluc} shows the additional estimation error due to fluctuations of the signals, characterized by the variance of the normal distribution.
Here, the additional estimation error is calculated by the difference between the average estimation error obtained by the Monte-Carlo simulation,
\begin{align}
\Delta^2 N_c\equiv \langle (N_c^\text{est}-N_c)^2\rangle,~~~\Delta^2 N_s\equiv \langle (N_s^\text{est}-N_s)^2\rangle
\end{align}
and the estimation error indicated by Fisher information.
Explicitly, the simulation is performed using the maximum likelihood estimation, and the estimation error from Fisher information is written as $1/MF(N_c)$ and $1/MF(N_s)$.
We can see from the figure that the amount of the fluctuation of the signals adds the estimation error.
It can be understood by noting that when fluctuations are small, Fisher information can be approximated by that at the mean so that the only the fluctuation changes the total estimation error.
Thus, when fluctuation is large as shown in Fig. \ref{fluc} (a), one observes the deviation because the Fisher information cannot be simply approximated by that at the mean.
In conclusion, assuming that the fluctuation of the signal strength is small enough, Fisher information at the mean is an essential quantity that determines the relevant estimation error.

\bibliography{reference}

\end{document}